# Quasi-logarithmic spacing law in dewetting patterns from the drying meniscus of a polymer solution

Yong-Jun Chen [a, b,]*, Kosuke Suzuki [a, c], Hitoshi Mahara [a], Tomohiko Yamaguchi [a,]*

[a] *Nanosystem Research Institute, National Institute of Advanced Industrial Science and Technology (AIST), 1-1-1 Higashi, Tsukuba, Ibaraki 305-8565, Japan*
[b] *Department of Physics, Shaoxing University, Shaoxing City, Zhejiang Province 312000, China*
[c] *Graduate School of Pure and Applied Science, University of Tsukuba, 1-1-1 Tennodai, Tsukuba, Ibaraki 305-8571, Japan*

**ABSTRACT:**

We report on a periodic precipitation pattern emerged from a drying meniscus via evaporation of a polystyrene solution in a Petri dish. It appeared a quasi-logarithmic spacing relation in the pattern as a result of stick-slip motion of the contact line towards the wall. A model based on the dynamics of the evaporating meniscus is proposed.

*Corresponding author. Tel: +81-29-861-9408 / Fax: +81-29-861-6236.
*E-mail address*: tomo.yamaguchi@aist.go.jp (T. Yamaguchi)
            yongjunchen@ni.aist.go.jp (Y. J. Chen)

Keywords:   Precipitation; dewetting; stick-slip motion; pattern; modeling

# 1. Introduction

Pattern formation by deposition of nonvolatile components via evaporation of solvent from a thin solution is known as a dewetting process. It has been recognized as a unique method for the surface treatment to fabricate a regular pattern [1-13], and a potential application of such self-organized patterns has been pointed out for constructing device-oriented structures [5]. When a droplet of such a solution is dried on a solid substrate, often left is a concentric precipitation pattern through the stick-slip motion of the contact line [13, 14]. Also reported is the gradient pattern under confined geometry [9], occasionally associated with patterns with complex morphology [3, 5, 13]. Concerning the gradient pattern formation, a concentric one has been demonstrated in a mesoscopic scale [11].

The mechanism of such periodical depositions in the dewetting process has been an issue of discussion. The Marangoni flow transfers the solute along the interface of the evaporating solution towards the vicinity of the contact line to induce the precipitation of nonvolatile solute (for example, polymers, nanoparticles, and carbon nanotubes) [14]. This precipitation may lead to the pining of the contact line, which determines the distribution of the pattern. The motion of the contact line is dominated by the antagonistic balance between the pinning force from the solid substrate and the depinning force due to the surface tensions [13-16].

Usually, a layer of a polymer solution leaves a thin film after evaporation [17]. When the concentration of the polymer solution is low enough, we obtain a so-called dewetting pattern. A thin layer of the solution starts to dewet from the center of the Petri dish when the thickness of the solution is less than the critical thickness for dewetting [18, 19]. After the early dewetting process, the liquid layer with a meniscus is left along the wall of the Petri dish. The succeeding dewetting process of the meniscus is the main issue of this article.

In the present paper, we demonstrate by laboratory experiments in a Petri dish a macroscopic formation of a gradient concentric pattern from a polymer solution, and we discuss the mechanism of this pattern formation based on the evaporation dynamics of the meniscus. We propose a model to explain the stick-slip motion of the evaporating meniscus and examine the validity of this model by numerical calculations.

## 2. Experimental

As a nonvolatile solute, we used polystyrene (PS) (Aldrich, average molecular weight: 35,000) which was dissolved in toluene (1.0 ~ 5.0 mg mL$^{-1}$). A Petri dish (diameter: 5 or 11 cm) was filled with the PS solution so that the depth of the solution was ~ 1 cm (Figure 1(a)). The Petri dish filled with the PS solution was set in a vacuum chamber (~ 0.65 kPa). After evaporation of toluene, a thin film of PS was left on the bottom of the dish. Near the wall of the dish, a gradient concentric pattern was found by naked eye. The pattern was observed under a laser scanning microscope (KEYENCE VK 9710). All experiments were performed at room temperature.

## 3. Results

During the evaporation of the solution, a meniscus was formed along the glass wall of the Petri dish. It took ~ 15 min for complete drying of the solution under the present experimental conditions.

Figure 1(b) demonstrates a typical precipitation pattern. The concentric rings are distributed within the range of 1 cm from the wall of the Petri dish. The distance between the neighboring rings is in the scale of millimeters.

Figures 2(a) and (b) are the local images of the precipitation pattern observed under laser scanning microscope in two different Petri dishes with different diameters. The wavelength $\lambda_n$ (given as the center-to-center distance between the adjacent stripes) is plotted in Figures 2(c) and (d) in different ways. Figure 2(c) is the plots of $\log_{10} \lambda_n$ against the number of the stripe $n$ counted from the wall of the Petri dish. The plots fall on the straight lines, and the spacing rule seems logarithmic in these gradient patterns. However, when $\lambda_n$ is re-plotted against the position of the $n$-th stripe ($x_n$), it deviates from a straight line slightly upwards with the distance from the wall of the Petri dish (Figure 2(d)).

**4. Discussion**

As shown in Figure 2, the dewetting process at the meniscus region produced a gradient pattern with a self-similar spacing relation that is also found in microscopic concentric dewetting patterns, Liesegang bands [20], and many biological systems [11]. If the deposition occurs exponentially, i.e., $x_n \propto \exp(2n\pi + \alpha)$, the wavelength, $\lambda_n = x_n - x_{n-1}$, should satisfy the following logarithmic relations: $\log \lambda_n = 2n\pi + \text{const.}$ and $\lambda_n \propto x_n$. The former one is satisfactory as shown in Figure 2(c), but the latter one tested in Figure 2(d) is not. These experimental results indicate that the wavelength does not strictly obey the logarithmic relation. In this sense, the spacing rule in the present experiments is not logarithmic but quasi-logarithmic.

In order to approach the origin of this quasi-logarithmic relation, we have constructed a model for describing the stick-slip motion of the evaporating meniscus. The outline of this model is as follows: (a) Evaporation increases the precipitation of the solute from the saturated solution to increase the pinning force (assumption). (b) Concomitantly, evaporation decreases the contact angle at the pinned contact line to increase the de-pinning force. (c) The contact line is de-pinned when the de-pinning force overcomes the pinning force and moves to the next pinning position. (d) The dynamic balance between these antagonistic forces is described through the time evolution of the contact angle.

*4.1. Shape of Meniscus*

The shape of the meniscus can be described by the following equation; see Appendix:

$$\frac{dh}{dx} = -\left\{\gamma_{a/l}^2 \left[P_{\text{dif}}(H)h + \frac{1}{2}\rho g h^2 - \gamma_{a/l}\cos\theta\right]^{-2} - 1\right\}^{1/2}, \qquad (1)$$

where $h$ and $x$ are the height at the surface of the meniscus (see Figure 3) and the displacement from the boundary of the Petri dish, respectively; $\rho$, $g$, $\gamma_{a/l}$ and $\theta$ are the density of the solution, the acceleration of gravity, the surface tension at the air-liquid interface, and the contact angle, respectively; $H$ is the maximum height of the meniscus, and $P_{dif}(H)$ stands for the maximum pressure difference:

$$P_{dif}(H) \equiv P_0 - P_{sub} = \frac{\gamma_{a/l}(\cos\theta - \sin\theta)}{H} - \frac{1}{2}\rho g H . \tag{2}$$

Below we consider the time evolution of the meniscus whose contact line is initially pinned at the position $L_n$ with the initial contact angle $\theta_0$, where $\theta = \theta(t) \leq \theta_0$.

*4.2. Pinning Force and Receding Angle*

The receding contact angle $\theta_r$ is known to be related to the roughness of the solid surface [19]. The receding contact angle in this article means the *initial receding contact angle* that is a critical value for the slip motion of contact line to occur [21]. According to the experimental results of Dettre and Johnson by using pure solvents [19, 22], the approximate profile of the receding contact angle against the roughness of the solid substrate is illustrated by two regimes, I and II, as shown in the inset in Figure 3. In regime II, no slip motion occurs because the receding contact angle increases with an increase in the roughness (i.e., the contact line is always pinned). Contrarily, the contact line may undergo the slip motion in the regime I when the de-pinning force is strong enough to overcome the pinning force. Although the relation between the receding contact angle and the roughness of the substrate is not well known yet, we may reasonably expect the following relation in regime I according to the trend of the experimental curve [19, 22]:

$$\theta_r \sim \theta_0 - R^\beta, \tag{3}$$

where $R$ is the roughness on the substrate and $\beta$ is an exponent.

We assume a similar relation in the present dewetting system. The roughness $R$ being replaced with the mass of deposition $M(t)$ on the substrate, we obtain

$$\theta_r(t) = \theta_0 - \chi(M(t))^\beta, \tag{4}$$

where $\chi$ is a constant. This equality holds as far as the pinning force is equal to or stronger than the depinning force. Thus, the pinning force is implicitly given in the present model by the receding contact angle $\theta_r(t)$ via $M(t)$. The value of $M(t)$ is determined by the volume of the solvent evaporated from the pinned meniscus, $V_{\text{vap}}(t)$:

$$M(t) = cV_{\text{vap}}(t) = cv\int\!\!\int_S \mathrm{d}s\mathrm{d}t \quad, \tag{5}$$

where $c$ is the concentration of saturation, $v$ is the rate of evaporation, $S$ is the surface area of the evaporating meniscus, and the time integration is started from the onset of pinning.

*4.3. De-Pinning Force*

Evaporation from the pinned meniscus decreases the contact angle and consequently increases the de-pinning force $F_{\text{dep}}(t)$:

$$F_{\text{dep}}(t) = \gamma_{a/l} \cos\theta(t) + \gamma_{l/s} \quad, \tag{6}$$

where $\gamma_{l/s}$ is the interfacial tension at the liquid-substrate interface, and $F_{\text{dep}}(t)$ is defined for the unit length.

*4.4. Onset of De-Pinning and Slipping*

When the de-pinning force exceeds the pinning force, the pinned contact line slips to the stable next position where the contact angle is reset to be equal to the initial contact angle $\theta_0$. This condition for the onset of de-pinning is given by

$$\theta(t) < \theta_r(t), \tag{7}$$

or

$$\arccos\left(\frac{F_{\text{dep}}(t) - \gamma_{l/s}}{\gamma_{a/l}}\right) < \theta_0 - \chi(M(t))^\beta, \tag{8}$$

by using Eqs. (4) and (6). In numerical calculations, $\theta(t)$ can is determined from the time evolution of the meniscus (see Eqs. (1) and (5)).

Assuming that the volume stays unchanged during the slipping process, we obtain Eq. (9) that determines the next pinning position $L_{n+1}$:

$$\left(\int_0^{L_n} h(x)\mathrm{d}x\right)_{t_n=\tau_n} = \left(\int_0^{L_{n+1}} h(x)\mathrm{d}x\right)_{t_{n+1}=0}, \tag{9}$$

where $h(x)$ is the height of the meniscus in Eq. (1); $t_n$ is the time in the $n$-th cycle and $\tau_n$ is its duration. Since $L_n$ and $\theta(t_n=\tau_n)$ on the left hand side and $\theta(t_{n+1}=0)=\theta_0$ on the right hand side are both given, the re-pinning position $L_{n+1}$ can be calculated.

In this way, the initial conditions are reset in every slipping process and the stick-slip motion proceeds under constant evaporation of the solvent.

*4.5. Numerical Calculations*

We implemented by numerical calculations the stick-slip motion of the meniscus based on the above-mentioned model. Figure 4 shows the results of simulation concerning the relation between the wavelength $\lambda_n = L_{n-1} - L_n$ and the position $L_n$. Figures 4(a), (b) and (c) illustrate the influence of the concentration of saturation ($c$), the rate of evaporation ($v$) and the exponent on the mass of deposition ($\beta$), respectively. The plots of $\log_{10} \lambda_n$ against $n$ appear similar to those in Figure 2(c) (data not shown).

As typically indicated in Figure 4(a), the wavelength is not linearly dependent on the pinning position. It suggests that the quasi-logarithmic relation observed in Figure 2(d) is not an experimental artifact but is due to the stick-slip motion of the meniscus, and that the dewetting pattern at the meniscus region is well implemented by the present model. Also noteworthy is the strong dependence of the wavelength on the concentration of saturation. Contrarily, Figure 4(b) indicates little influence of evaporation rate on the pattern morphology. A similar tendency in laboratory experiments has been reported [23, 24]. Figure 4(c) shows the sensitivity of the wavelength to the value of exponent $\beta$. A large value of $\beta$ corresponds to the remarkable effect of the deposited mass to the receding angle $\theta_r(t)$, which results in the decrease in the wavelength as indicated. The effects of these three parameters are tightly related to each other in laboratory experiments through the nature of the solvents. In this sense, a detailed study by changing the solvent remains as a future challenge.

Our simulation based on the proposed model has qualitatively reproduced the quasi-logarithmic spacing relation in the dewetting pattern. The present work suggests that the quasi-logarithmic spacing is not an artifact in laboratory experiments but is the characteristic nature of dewetting patterns at the meniscus region. It also supports that our assumption introduced in Eq. (4), the equivalence of the role of the deposition mass to the roughness of the substrate, is phenomenologically reasonable. The origin of quasi-logarithmic spacing is not clear at this moment, but we tend to ascribe it to the influence of gravity.

**5. Conclusions**

We have demonstrated that a macroscopic precipitation pattern of polystyrene can be produced by evaporation of solvent at the meniscus region. The stick-slip motion of the meniscus is modeled by taking into account the antagonistic balance between the pinning force and the de-pinning force. The origin of the pinning force is the mass deposition which phenomenologically corresponds to the roughness of the solid substrate, and the origin of the de-pinning force is the contact angle and the interfacial tensions at the liquid surface. Our simulation has reproduced the quasi-logarithmic spacing relation observed in laboratory experiments, which suggests that the stick-slip motion depends on the concentration of the solution and the mass of deposition on the substrate. Although the precipitation pattern resembles that of Liesegang patterns [20], the mechanism and the resulting quasi-logarithmic spacing relation are both decisively different in the dewetting gradient patterns at the meniscus region.

**Acknowledgements**

We thank Drs. Yusuke Hara and Takashi Iwatsubo for their kind help for the experiments, and Dr. Tetsuya Yamamoto for useful comments. This work was supported by the Japanese Ministry of Education, Culture, Sports, Science and Technology (MEXT) via Grant-in-Aid for Scientific Research on Innovative Areas "Emergence in Chemistry" (Grant No. 20111007).

**Appendix: The shape of meniscus**

We consider an equilibrium meniscus shown in Figure 3, where $x$ and $h$ are the orthogonal coordinates. The pressure balance on the air-liquid surface of the meniscus can be expressed as [25]

$$P_0 + \gamma_{a/l}(\kappa_1 + \kappa_2) = P_{sub} - \rho g h, \tag{A1}$$

where $P_0$, $P_{sub}$, $\gamma_{a/l}$, $\rho$, $h$, and $g$ are the pressure in the chamber, the pressure at the substrate, the surface tension of air-liquid interface, the density of the solution, the height of the meniscus, and the acceleration of gravity, respectively; $\kappa_1$ and $\kappa_2$ are the principal curvatures of meniscus surface. Since the meniscus can be approximated as a part of a cylinder surface when the diameter of the Petri dish is large enough, we set

$$\kappa_1 = -h''/(1+h'^2)^{-3/2} \tag{A2}$$

and

$$\kappa_2 = 0. \tag{A3}$$

Thus,

$$P_0 - P_{sub} + \rho g h = h''/(1+h'^2)^{-3/2}, \tag{A4}$$

or, by setting $y = h'^2$, we obtain

$$P_0 - P_{sub} + \rho g h = \frac{\gamma_{a/l}}{2(1+y)^{3/2}} \frac{dy}{dh}. \tag{A5}$$

At the bottom of the Petri dish ($h=0$), $y = \tan^2 \theta$ ($\theta$ is the contact angle between the solution and the dry substrate). At the wall of the Petri dish ($h = H$, where $H$ is the maximum height of the meniscus), $y = \tan^2(\pi/2 - \theta)$. Here, we assume that the contact angles are the same on the wall and on the bottom. Using these boundary conditions and setting

$$P_{dif}(H) \equiv P_0 - P_{sub} = \frac{\gamma_{a/l}(\cos\theta - \sin\theta)}{H} - \frac{1}{2}\rho g H, \tag{A6}$$

we get the following equation by integration of Eq. (A5):

$$\frac{dh}{dx} = -\left\{ \gamma^2 [P_{dif}(H)h + \frac{1}{2}\rho g h^2 - \gamma_{a/l}\cos\theta]^{-2} - 1 \right\}^{1/2}. \tag{A7}$$

This equation, which appears as Eq. (1) in the text, determines the shape of the meniscus in the present system.

**Figure captions**

**Figure 1.** Experimental setup and concentric precipitation pattern in macroscopic scale. (a) Experimental setup. (b) A typical concentric pattern in a Petri dish. The initial concentration of polystyrene (PS) in toluene was 1.0 mg mL$^{-1}$.

**Figure 2.** Gradient precipitation patterns in two different dishes with different diameters. (a) Stripes in a big Petri dish (diameter: 11 cm). (b) Stripes in a small Petri dish (diameter: 5 cm). Scale bar: 1 mm. (c) Plots of $\log_{10} \lambda_n$ against the order of deposition $n$, where $\lambda_n$ is the wavelength ($\lambda_n = x_n - x_{n-1}$) in mm and $x_n$ is the position of the $n$-th stripe from the boundary of dish. The dashed line is the curve fitting to the data. (d) Wavelength against $x_n$. The concentration of PS solution is 1 mg mL$^{-1}$. Molecular weight of PS is 35000.

**Figure 3.** Schematic illustration of the meniscus in the Petri dish. The inset illustrates the approximate curve of the receding contact angle against the roughness of the substrate after Refs. [19, 22].

**Figure 4.** Wavelength in the simulated stick-slip motion. The position at $L_n$ = 0 corresponds to the position of the wall. (a) Influence of concentration of the solution. Parameters for simulation: $v$ = 0.001 and $\beta = 0.50$. (b) Influence of the rate of evaporation. Parameters for simulation: $c$ = 6.0 and $\beta = 0.50$. (c) Influence of exponent $\beta$. Parameters for simulation: $c$ = 6.0 and $v = 0.001$. Common parameters: $\chi = 2000.0$, $g = 10.0$, $\gamma_{a/l} = 0.05$, $\rho = 0.1$, and $\theta_0 = 2\pi/9$. The dashed lines are the curve-fittings to the data.

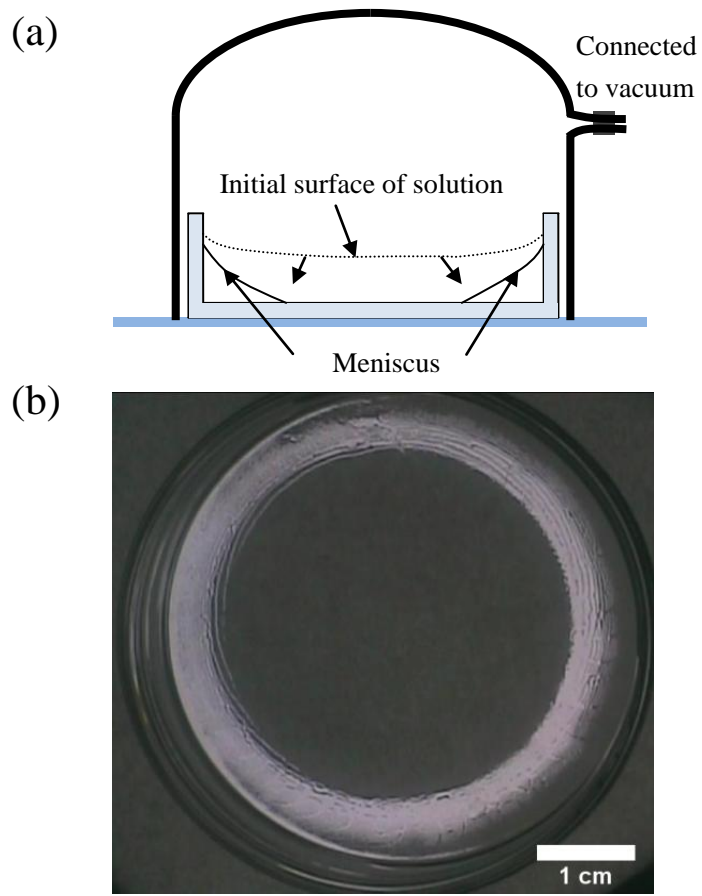

Figure 1: Yong-Jun Chen, Kosuke Suzuki, Hitoshi Mahara, Tomohiko Yamaguchi

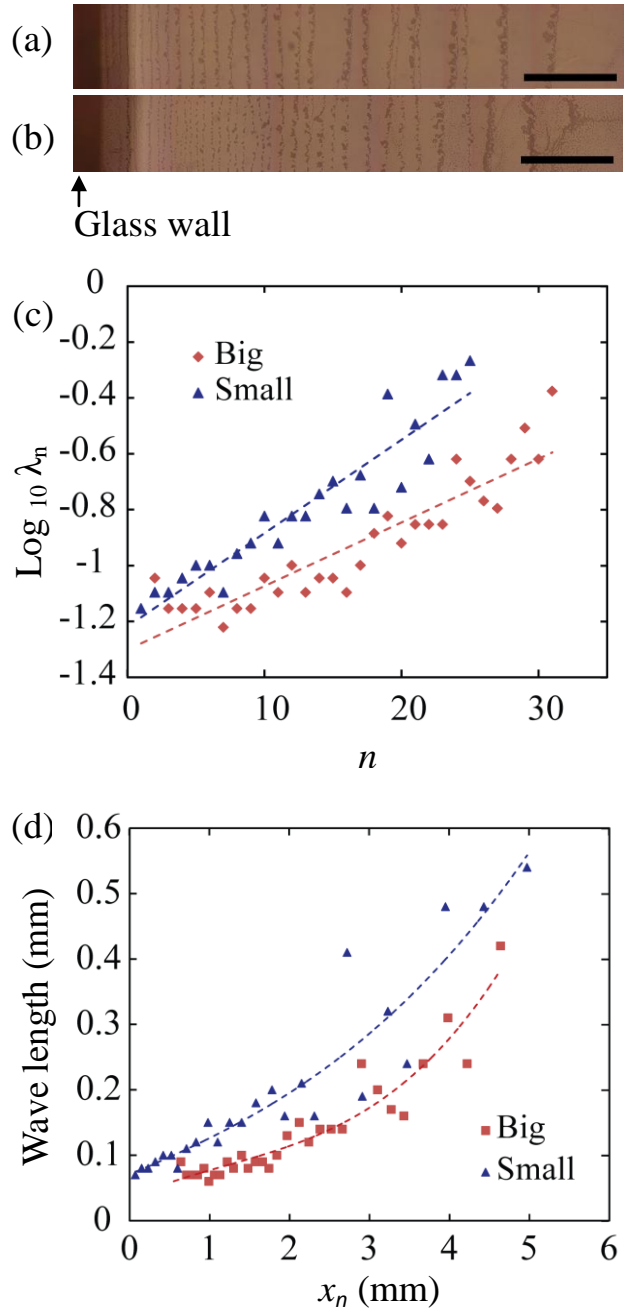

Figure 2: Yong-Jun Chen, Kosuke Suzuki, Hitoshi Mahara, Tomohiko Yamaguchi

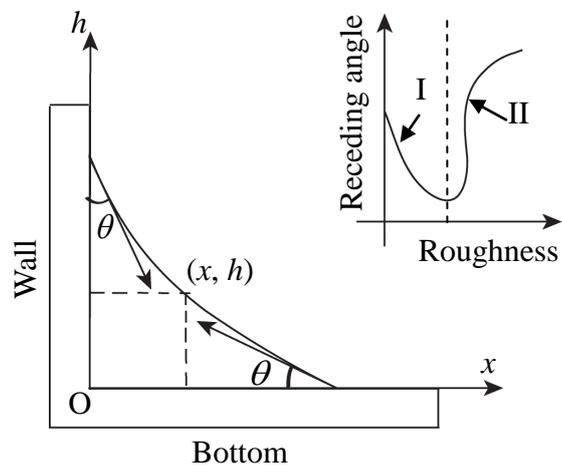

Figure 3: Yong-Jun Chen, Kosuke Suzuki, Hitoshi Mahara, Tomohiko Yamaguchi

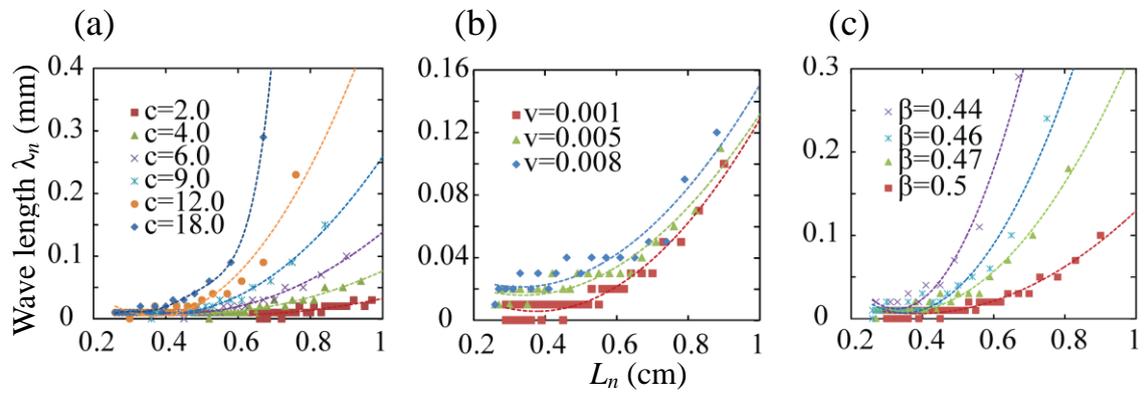

Figure 4 : Yong-Jun Chen, Kosuke Suzuki, Hitoshi Mahara, Tomohiko Yamaguchi